\documentclass[aps,pra,twocolumn,amsfonts,amssymb,amsmath,showpacs,
floatfix,nofootinbib,groupedaddress,superscriptaddress,citesort]{revtex4}
\usepackage{mathrsfs}
\usepackage{amsfonts}
\usepackage{amstext}
\usepackage{amsmath}
\usepackage{amssymb}
\usepackage{bm}
\usepackage{bbm}
\usepackage[dvips]{graphicx}
\def\qed{\leavevmode\unskip\penalty9999 \hbox{}\nobreak\hfill
     \quad\hbox{\leavevmode  \hbox to.77778em{%
              \hfil\vrule   \vbox to.675em%
               {\hrule width.6em\vfil\hrule}\vrule\hfil}}
     \par\vskip3pt}

\begin{document}

\title{The Wigner-Yanase  information can increase under phase sensitive incoherent operations }

\author{Shuanping Du}
\email{shuanpingdu@yahoo.com} \affiliation{School of Mathematical
Sciences, Xiamen University, Xiamen, Fujian, 361000, China}

\author{Zhaofang Bai}\thanks{Corresponding author}
\email{baizhaofang@xmu.edu.cn} \affiliation{School of Mathematical
Sciences, Xiamen University, Xiamen, Fujian, 361000, China}

\begin{abstract}

We found that the Wigner-Yanase skew information, which has been
recently proposed as a measure of coherence in [Phys. Rev. Lett.
\textbf{113}, 170401(2014)], can increase under a class of
operations which may be interpreted as incoherent following the
framework of Baumgratz et al., while being phase sensitive.



\end{abstract}

\maketitle

{\it Introduction---}Coherence arising from quantum superposition
plays a central role for quantum mechanics. Quantum coherence is
an important subject in quantum theory and quantum information
science which is a common necessary condition for both
entanglement and other types of quantum correlations. It is the
key resource for quantum technology, with applications in quantum
optics, information processing, metrology and cryptography. Up to
now, several themes of coherence have been considered such as
witness coherence \cite{Li}, catalytic coherence \cite{Abe}, the
thermodynamics quantum coherence \cite{Ces}, and the role of
coherence in biological system \cite{Hue}. But, given a quantum
state, how much the coherence does it contain? How to quantify the
quantum coherence? There is no well-accepted efficient method to
quantify the coherence in quantum system until recently. T.
Baumgratz etc. have introduced a rigorous framework for the
quantification of quantum coherence \cite{BCP}. Let ${\mathcal H}$
be a finite dimensional Hilbert space with $d=\dim({\mathcal H})$.
Fixing a basis $\{|i\rangle\}_{i=1}^d$, we call all density
operators (quantum states) that are diagonal in this basis
incoherent, and this set of quantum states will be labelled by
${\mathcal I}$. Quantum operations are specified by a finite set
of Kraus operators $\{A_n\}$ satisfying $\sum_n A_n^\dag A_n=I$,
$I$ is the identity operator on ${\mathcal H}$. From \cite{BCP},
quantum operations are incoherent if they fulfil $A_n\rho
A_n^\dag/Tr(A_n\rho A_n^\dag)\in {\mathcal I}$ for all $\rho\in
{\mathcal I}$ and for all $n$. Based on Bumgratz's suggestion
\cite{BCP}, any proper measure of coherence ${\mathcal C}$ must
satisfy the following axiomatic postulates.

(i) The coherence vanishes on the set of incoherent states
(faithful criterion), ${\mathcal C}(\rho)=0$ for all
$\rho\in{\mathcal I}$;

(ii) Monotonicity under incoherent operations $\Phi$, ${\mathcal
C}(\Phi(\rho))\leq {\mathcal C}(\rho)$;

(iii) Non-increasing under mixing of quantum states (convexity),
$${\mathcal C}(\sum_np_n\rho_n)\leq \sum_np_n{\mathcal
C}(\rho_n)$$ for any ensemble $\{p_n,\rho_n\}$.

For every self-adjoint non-degenerate $d\times d$ diagonal matrix
$K$, it was shown that the Wigner-Yanase skew information
$${\mathcal C}(\rho, K)=-\frac{1}{2}Tr([\sqrt{\rho},K]^2)$$ satisfies the above three postulates and so is a
measure of the $K$-coherence of the state $\rho$ \cite{Gir}, where
$[\sqrt{\rho},K]=\sqrt{\rho}K-K\sqrt{\rho}$.\vspace{0.1in}
However, the following example shows that the Wigner-Yanase skew
information can increase under an   incoherent operation.

Assume $\dim({\mathcal H})=3$. Let $$\begin{array} {ll}
K&=|1\rangle\langle 1|+10|2\rangle\langle
2|+5|3\rangle\langle 3|,\\
|\psi\rangle&=\frac{1}{\sqrt 3}|1\rangle+\frac{1}{\sqrt
3}|2\rangle+\frac{1}{\sqrt 3}|3\rangle,\\
|\phi\rangle&=\frac{1}{\sqrt 2}|1\rangle+\frac{1}{\sqrt
2}|2\rangle,\\ A_1&=\frac{1}{\sqrt 2}|1\rangle\langle
1|+\frac{1}{\sqrt 2}|2\rangle\langle
2|,\\
A_2&=\frac{1}{\sqrt 2}|1\rangle\langle 2|+\frac{1}{\sqrt
2}|2\rangle\langle
3|,\\
A_3&=\frac{1}{\sqrt 2}|1\rangle\langle 3|+\frac{1}{\sqrt
2}|2\rangle\langle 1|.\end{array}$$ It is easy to see that
$\sum_{n=1}^3 A_n^\dag A_n=I$. Furthermore, for any diagonal
density operator $\rho\in {\mathcal I}$, we have $A_n\rho
A_n^\dag/Tr(A_n\rho A_n^\dag)\in {\mathcal I}$ for all $n$. That
is to say, the Kraus operators $\{A_n\}_{n=1}^3$ define an
incoherent operation. By a direct computation, one can get
$A_n|\psi\rangle=\frac{1}{\sqrt 3}|\phi\rangle, n=1,2,3$. Hence
$$\sum_{n=1}^3A_n|\psi\rangle\langle\psi|A_n^\dag=|\phi\rangle\langle\phi|.$$
However, it is easy to check that $${\mathcal
C}(|\phi\rangle\langle \phi|,K)=\frac{81}{4}> {\mathcal
C}(|\psi\rangle\langle \psi|,K)=\frac{122}{9}.$$ This shows that
Wigner-Yanase skew information ${\mathcal C}(\cdot, K) $ can
increase under incoherent operations.

In this paper, it will be  proved that the Wigner-Yanase skew
information can increase under a class of phase sensitive
incoherent operations.


{\it Result---For any self-adjoint non-degenerate $d\times d
(d\geq 3)$ diagonal matrix $K$, ${\mathcal C}(\cdot, K) $ can
increase under a class of phase sensitive incoherent operations.}

{\bf Proof.} The main idea is to pick a representative of a
peculiar class of operations, incoherent but phase sensitive and
show that the skew information increases. We firstly treat the
case $d=3$. Because $K$ is non-degenerate, we have $\lambda_i\neq
0$ and $\lambda_i\neq\lambda_j, i,j=1,2,3$. The proof is divided
into several cases.

{\bf Case i.} $\lambda_1<\lambda_3<\lambda_2$.

Let $$\begin{array} {ll} |\psi\rangle&=\frac{1}{\sqrt
3}|1\rangle+\frac{1}{\sqrt 3}|2\rangle+\frac{1}{\sqrt
3}|3\rangle,\\
|\phi\rangle&=\frac{1}{\sqrt
2}|1\rangle+\frac{1}{\sqrt 3}|2\rangle+\frac{1}{\sqrt
6}|3\rangle,\\
A_1&=\frac{1}{\sqrt 2}|1\rangle\langle 1|+\frac{1}{\sqrt
3}|2\rangle\langle
2|+\frac{1}{\sqrt 6}|3\rangle\langle 3|,\\
A_2&=\frac{1}{\sqrt 2}|1\rangle\langle 2|+\frac{1}{\sqrt
3}|2\rangle\langle 3|+\frac{1}{\sqrt 6}|3\rangle\langle
1|,\\
A_3&=\frac{1}{\sqrt 2}|1\rangle\langle 3|+\frac{1}{\sqrt
3}|2\rangle\langle 1|+\frac{1}{\sqrt 6}|3\rangle\langle
2|.\end{array}$$ By a direct computation, we have $\sum_{n=1}^3
A_n^\dag A_n=I$. Moreover, for any diagonal density operator
$\rho\in {\mathcal I}$,  $A_n\rho A_n^\dag/Tr(A_n\rho A_n^\dag)\in
{\mathcal I}$ for all $n$. That is, the Kraus operators
$\{A_n\}_{n=1}^3$ define an incoherent operation. It is easy to
check that $A_n|\psi\rangle=\frac{1}{\sqrt 3}|\phi\rangle,
n=1,2,3$. Therefore
$$\sum_{n=1}^3A_n|\psi\rangle\langle\psi|A_n^\dag=|\phi\rangle\langle\phi|.$$
After simple computation, we obtain
$$\begin{array}{ll} {\mathcal C}(|\phi\rangle\langle
\phi|,K)-{\mathcal C}(|\psi\rangle\langle
\psi|,K)\\
=\frac{(\lambda_3-\lambda_1)(3\lambda_2-3\lambda_3+\lambda_2-\lambda_1)}{36}>0.\end{array}$$
Therefore Wigner-Yanase skew information ${\mathcal C}(\cdot, K) $
increases under incoherent operations.

{\bf Case ii.} $\lambda_2<\lambda_3<\lambda_1$.

Let $$\begin{array} {ll} |\psi\rangle&=\frac{1}{\sqrt
3}|1\rangle+\frac{1}{\sqrt 3}|2\rangle+\frac{1}{\sqrt
3}|3\rangle,\\
|\phi\rangle&=\frac{1}{\sqrt 3}|1\rangle+\frac{1}{\sqrt
2}|2\rangle+\frac{1}{\sqrt
6}|3\rangle,\\
A_1&=\frac{1}{\sqrt 3}|1\rangle\langle 1|+\frac{1}{\sqrt
2}|2\rangle\langle
2|+\frac{1}{\sqrt 6}|3\rangle\langle 3|,\\
A_2&=\frac{1}{\sqrt 3}|1\rangle\langle 2|+\frac{1}{\sqrt
2}|2\rangle\langle 3|+\frac{1}{\sqrt 6}|3\rangle\langle
1|,\\
A_3&=\frac{1}{\sqrt 3}|1\rangle\langle 3|+\frac{1}{\sqrt
2}|2\rangle\langle 1|+\frac{1}{\sqrt 6}|3\rangle\langle
2|.\end{array}$$ It is easy to see that $\sum_{n=1}^3 A_n^\dag
A_n=I$. After simple computation, for any diagonal density
operator $\rho\in {\mathcal I}$, we have $A_n\rho
A_n^\dag/Tr(A_n\rho A_n^\dag)\in {\mathcal I}$ for all $n$. That
is to say, the Kraus operators $\{A_n\}_{n=1}^3$ define an
incoherent operation. By a direct computation,
$A_n|\psi\rangle=\frac{1}{\sqrt 3}|\phi\rangle, n=1,2,3$. Hence
$$\sum_{n=1}^3A_n|\psi\rangle\langle\psi|A_n^\dag=|\phi\rangle\langle\phi|.$$ However, it is easy to check
$$\begin{array}{ll} {\mathcal C}(|\phi\rangle\langle
\phi|,K)-{\mathcal C}(|\psi\rangle\langle
\psi|,K)\\
=\frac{(\lambda_3-\lambda_2)(3\lambda_1-3\lambda_3+\lambda_1-\lambda_2)}{36}>0.\end{array}$$
This tells us that Wigner-Yanase skew information ${\mathcal
C}(\cdot, K) $ increases.

For the remained cases, in order to construct suitable pure state
$|\phi\rangle$, one only need to adjust the position of
$\frac{1}{\sqrt 2},\frac{1}{\sqrt 3},\frac{1}{\sqrt 6}$ according
to the order relation of $\lambda_1,\lambda_2,\lambda_3$. For
instance, in the case of $\lambda_3<\lambda_1<\lambda_2$, we
choose
$$|\phi\rangle=\frac{1}{\sqrt 6}|1\rangle+\frac{1}{\sqrt
3}|2\rangle+\frac{1}{\sqrt 2}|3\rangle.$$ The incoherent Kraus
operators are constructed as follows, $$\begin{array} {ll}
A_1&=\frac{1}{\sqrt 6}|1\rangle\langle 1|+\frac{1}{\sqrt
3}|2\rangle\langle
2|+\frac{1}{\sqrt 2}|3\rangle\langle 3|,\\
A_2&=\frac{1}{\sqrt 6}|1\rangle\langle 2|+\frac{1}{\sqrt
3}|2\rangle\langle 3|+\frac{1}{\sqrt 2}|3\rangle\langle
1|,\\
A_3&=\frac{1}{\sqrt 6}|1\rangle\langle 3|+\frac{1}{\sqrt
3}|2\rangle\langle 1|+\frac{1}{\sqrt 2}|3\rangle\langle
2|.\end{array}$$

For the general case $\dim {\mathcal H}=d$,
$$|\psi\rangle=\frac{1}{\sqrt d}|1\rangle+\frac{1}{\sqrt
d}|2\rangle+\cdots+\frac{1}{\sqrt d}|d\rangle$$ and
$$K=\lambda_1|1\rangle\langle 1|+\cdots+\lambda_d|d\rangle\langle
d|(\lambda_i\neq\lambda_j,i,j=1,2,\cdots,d).$$ The construction of
pure states and incoherent Kraus operators is originated from the
case $\dim{\mathcal H}=3$. Assume
$\lambda_1<\lambda_2<\lambda_3<\cdots<\lambda_d$. Let
$$\begin{array}{ll}
|\phi\rangle&=\sqrt{\frac{3}{2d}}|1\rangle+\sqrt{\frac{1}{2d}}|2\rangle+\sqrt{\frac{1}{d}}|3\rangle+\cdots+
\sqrt{\frac{1}{d}}|d\rangle.\\
\end{array}$$ By a direct computation, we have
$$\begin{array}{ll}
&C(|\phi\rangle\langle\phi|,K)-C(|\psi\rangle\langle\psi|,K)\\
&=\frac{\lambda_2-\lambda_1}{4d^2}[4\sum_{i=3}^d\lambda_i-(2d-5)\lambda_1-(2d-3)\lambda_2]\\
&>0.\end{array}$$ Let $$\begin{array} {ll}
A_1&=\sqrt{\frac{3}{2d}}|1\rangle\langle
1|+\sqrt{\frac{1}{2d}}|2\rangle\langle
2|+\sqrt{\frac{1}{d}}|3\rangle\langle
3|+\cdots\\
&+\sqrt{\frac{1}{d}}|d\rangle\langle d|,\\
A_2&=\sqrt{\frac{3}{2d}}|1\rangle\langle
2|+\sqrt{\frac{1}{2d}}|2\rangle\langle
3|+\sqrt{\frac{1}{d}}|3\rangle\langle 4|+\cdots\\
&+\sqrt{\frac{1}{d}}|d-1\rangle\langle d|+\sqrt{\frac{1}{d}}|d\rangle\langle 1|,\\
 A_3&=\sqrt{\frac{3}{2d}}|1\rangle\langle
3|+\sqrt{\frac{1}{2d}}|2\rangle\langle
4|+\sqrt{\frac{1}{d}}|3\rangle\langle
5|+\cdots\\
&+\sqrt{\frac{1}{d}}|d-1\rangle\langle
1|+\sqrt{\frac{1}{d}}|d\rangle\langle
2|,\\
\cdots & \\
A_i=&\sqrt{\frac{3}{2d}}|1\rangle\langle
m_i|+\sqrt{\frac{1}{2d}}|2\rangle\langle
m_{i+1}|+\sqrt{\frac{1}{d}}|3\rangle\langle
m_{i+2}|+\cdots\\
&+\sqrt{\frac{1}{d}}|s\rangle\langle m_{s+i-1}|+\cdots
+\sqrt{\frac{1}{d}}|d\rangle\langle
m_{m_{d+i-1}}|,\\
\cdots & \\
A_d&=\sqrt{\frac{3}{2d}}|1\rangle\langle
d|+\sqrt{\frac{1}{2d}}|2\rangle\langle
1|+\cdots+\sqrt{\frac{1}{d}}|d\rangle\langle d-1|,\end{array}$$
here $m_x=x-[\frac{x-1}{d}]d$, $[\cdot]$ is the greatest integer
function. It is easy to see that the Kraus operators
$\{A_i\}_{i=1}^d$ define an incoherent operation. Moreover,
$\sum_{i=1}^dA_i|\psi\rangle\langle\psi|
A_i^\dag=|\phi\rangle\langle\phi|$.


Generally, we can set $\lambda_{i_0}, \lambda_{j_0},\lambda_{k_0}$
such that $\lambda_{i_0}<\lambda_{j_0}<\lambda_{k_0}<\lambda_k,
k\neq i_0,j_0,k_0$. Without loss of generality, suppose
$i_0<j_0<k_0$. Let $\begin{array}{ll}
|\phi\rangle=&\sqrt{\frac{1}{d}}|1\rangle+\cdots+\sqrt{\frac{1}{d}}|i_0-1\rangle+\sqrt{\frac{3}{2d}}|i_0\rangle+\sqrt{\frac{1}{d}}|i_0+1\rangle\\
&+\cdots+\sqrt{\frac{1}{d}}|j_0-1\rangle+\sqrt{\frac{1}{2d}}|j_0\rangle+\sqrt{\frac{1}{d}}|j_0+1\rangle+\cdots\\
&+\sqrt{\frac{1}{d}}|k_0\rangle+\cdots+\sqrt{\frac{1}{d}}|d\rangle.\end{array}$
Similarly, one can construct a desired incoherent operation.
$\Box$

{\it Conclusion---}We have shown that the Wigner-Yanase skew
information can increase under a class of phase-sensitive
incoherent operations.  It is concluded that the Wigner-Yanase
skew information does not satisfy the framework of Baumgratz et
al. for measure of coherence.

The Wigner-Yanase skew information is actually a good measure of
asymmetry \cite{IR}. Recall that asymmetry is the properties of
states to be sensitive to phase shifts along some direction. An
asymmetry measure quantifies how much the symmetry in question is
broken by a given state and is widely investigated in recent years
\cite{IR,Bart,Gour1,Vacc,Gour2,Tolo1,Tolo2,Mar2}. States that are
asymmetric must have coherence and a non-trivial asymmetry measure
must be able to detect such coherence \cite{IR}. The author of
\cite{Gir} just does not distinguish between coherence and
asymmetry, arguing that asymmetry is coherence in a specific
basis. It is matter of taste to decide if the set of phase
intensive operations or the set of incoherent operations should be
considered the best pick for defining what is coherence.

{\it Acknowledgement---}The authors are indebt to referees for
their valuable comments. In particular, one of referees provides
\cite{IR} about asymmetry and its measure.

This work was completed while the authors were visiting the
Department of Mathematics and Statistics of the University of
Guelph during the academic year 2014-2015 under the support of
China Scholarship Council. We thank Professor David W. Kribs and
Professor Bei Zeng for their hospitality. This work is partially
supported by the Natural Science Foundation of China (No.
11001230), and the Natural Science Foundation of Fujian
(2013J01022, 2014J01024).



\end{document}